\documentclass{amsart}

\usepackage{amsmath,amssymb}
\usepackage{amsthm}
\usepackage[nobysame]{amsrefs}
\usepackage{indentfirst}
\usepackage{mathrsfs}
\usepackage{graphicx}
\usepackage{hyperref}
\usepackage{xcolor}
\usepackage{fourier}
\usepackage[margin=1.5in]{geometry}
\usepackage[norelsize,boxed,linesnumbered,vlined,algo2e]{algorithm2e} 
\usepackage{booktabs}
\usepackage{multirow}

\theoremstyle{plain}

\theoremstyle{definition}

\theoremstyle{remark}

\newcommand{\ud}{\,\mathrm{d}}

\newcommand{\TT}{\mathbb{T}}

\newcommand{\trans}{\top}

\newcommand{\wt}[1]{\widetilde{#1}}

\IfFileExists{mathabx.sty}%
  {\DeclareFontFamily{U}{mathx}{\hyphenchar\font45}%
   \DeclareFontShape{U}{mathx}{m}{n}{<->mathx10}{}%
   \DeclareSymbolFont{mathx}{U}{mathx}{m}{n}%
   \DeclareFontSubstitution{U}{mathx}{m}{n}%
   \DeclareMathAccent{\widebar}{0}{mathx}{"73}%
}{%
  \PackageWarning{mathabx}{%
    Package mathabx not available, therefore\MessageBreak substituting
    widebar with overline\MessageBreak }%
  \newcommand{\widebar}[1]{\overline{#1}}%
}
\newcommand{\wb}[1]{\widebar{#1}}

\newcommand{\eps}{\epsilon}

\newcommand{\Norm}[1]{\left\lVert#1\right\rVert}

\newcommand{\bra}[1]{\langle#1\rvert}

\newcommand{\ket}[1]{\lvert#1\rangle}

\newcommand{\braket}[2]{\langle#1\mid#2\rangle}

\newcommand{\grid}{\mathrm{grid}}
\newcommand{\aux}{\mathrm{aux}}

\title[A cubic scaling algorithm for excited states calculations in pp-RPA]
{A cubic scaling algorithm for excited states calculations in
  particle-particle random phase approximation}

\author{Jianfeng Lu} \address{Department of Mathematics, Department of
  Physics, and Department of Chemistry, Duke University, Box 90320, Durham NC 27708, USA}
\email{jianfeng@math.duke.edu}

\author{Haizhao Yang}
\address{Department of Mathematics, Duke University, Box 90320, Durham NC 27708, USA}
\email{haizhao@math.duke.edu}

\date{\today} \thanks{This work is partially supported by the National
  Science Foundation under grants DMS-1454939 and
  ACI-1450280. H.Y. thanks the support of the AMS-Simons Travel
  Award. We would like to thank Weitao Yang for helpful discussions.}

\begin{document}

\begin{abstract}
  The particle-particle random phase approximation (pp-RPA) has been
  shown to be capable of describing double, Rydberg, and charge
  transfer excitations, for which the conventional time-dependent
  density functional theory (TDDFT) might not be suitable. It is thus
  desirable to reduce the computational cost of pp-RPA so that it can
  be efficiently applied to larger molecules and even solids.  This
  paper introduces an $O(N^3)$ algorithm, where $N$ is the number of
  orbitals, based on an interpolative separable density fitting
  technique and the Jacobi-Davidson eigensolver to calculate a few
  low-lying excitations in the pp-RPA framework. The size of the
  pp-RPA matrix can also be reduced by keeping only a small portion of
  orbitals with orbital energy close to the Fermi energy. This reduced
  system leads to a smaller prefactor of the cubic scaling algorithm,
  while keeping the accuracy for the low-lying excitation energies.
\end{abstract}

\keywords{Excited states, particle-particle random phase
  approximation, density fitting, Jacobi-Davidson eigensolver}

\maketitle

\section{Introduction}
While the time-dependent density functional theory (TDDFT)
\cites{TDDFT,Runge1984} has been widely used in the prediction of
electronic excited states in large systems because of its low
computational cost and satisfying accuracy, it is known however that
TDDFT is not able to well describe double, Rydberg, charge transfer,
and extended $\pi$-systems excitations \cite{Dreuw2005}, which limits
its applications in many practical problems. This motivates the
development of the particle-particle random phase approximation
(pp-RPA) \cites{Aggelen2013,Peng2013,Scuseria2013} for excited state
calculations. It has been shown that the pp-RPA gives quite accurate
prediction of electronic excited states in moderate size molecular
systems~\cites{PengYang:2014,YangYang:2014}.

However, the application of the pp-RPA is still limited to small size
systems due to its expensive computational cost. Suppose $N$ is the
size of a given Hamiltonian after discretization, a naive
implementation takes $O(N^6)$ operations to solve the pp-RPA equation,
where $N$ is the number of orbitals. Recently, \cite{YangYang:2014}
proposed an $O(N^4)$ algorithm that is comparable with other commonly
used methods, e.g., configuration interaction singles (CIS) and TDDFT
methods.  To make the application of the pp-RPA feasible to larger
systems, this paper proposes an
$O(N N_{\text{aux}}^2+N^2N_{\text{aux}}+ N^2 N_{\text{grid}})$
algorithm based on a newly developed technique, the interpolative
separable density fitting in \cites{Lu2015,Lu2015Preprint}. Here
$N_{\aux}$ is the number of auxiliary basis functions used in the
density fitting and $N_{\grid}$ is the total number of real space grid
points, both scale linearly with $N$, and hence the overall scaling of
the proposed algorithm is $O(N^3)$.

In the numerical linear algebra point of view, the excited states
calculation in pp-RPA amounts to solving a generalized eigenvalue
problem. When focusing on low-lying excitations, the smallest (in
terms of the magnitude) few eigenpairs are desired. We refer the
readers to \cite{PengYang:2014} for the formal derivation of the
pp-RPA theory.

To simplify the discussion, let us consider systems in the domain with
periodic boundary condition, and without loss of generality, assumed
to be $\TT = [0,1]^d$ . After discretization (such as the
pseudo-spectral method employed in our numerical examples), 
the number of total spatial grid points is denoted by $N_{\text{grid}}$. Thus the
Hamiltonian operator $H$ becomes an $N_{\grid} \times N_{\grid}$ real
symmetric matrix. $\{(\epsilon_p,\phi_p)\}_{p=1,\dots, N_{\grid}}$ denote the $N_{\text{grid}}$
eigenpairs of $H$:
\begin{equation}
  H \phi_p = \eps_p \phi_p, \qquad \forall\, p = 1, \ldots, N_{\grid}. 
\end{equation}
The eigenvectors $\phi_p$ will be referred as orbitals and the
associated eigenvalues as orbital energy.  According to the Pauli's
exclusion principle, the low-lying eigenstates are occupied. 
The number of occupied orbitals is denoted by $N_{\text{occ}}$ (throughout this
work, we assume that the $N_{\text{occ}}$-th eigenvalue is
non-degenerate, \textit{i.e.},
$\eps_{N_{\text{occ}}} < \eps_{N_{\text{occ}} + 1}$). The rest of the
orbitals are virtual ones (also known as unoccupied orbitals).  The
virtual orbitals have higher orbital energy than the occupied ones;
the eigenvalues are separated by the Fermi energy:
\begin{equation}
  \eps_F = \frac{1}{2} \bigl( \eps_{N_{\text{occ}}} + \eps_{N_{\text{occ}} + 1} \bigr). 
\end{equation}
Therefore, the occupied orbitals have energy less than the Fermi energy while the virtual ones have energy higher than $\eps_F$. 

We follow the convention of quantum chemistry literature to use indices $i$,
$j$, $k$, and $l$ to index occupied orbitals, $a$, $b$, $c$, and $d$
for virtual orbitals, and $p$, $q$, $r$, and $s$ for unspecified
orbitals.  Assume that we consider the first $N_{\text{vir}}$ virtual
orbitals (ordered by eigenvalues) and
$N = N_{\text{occ}} + N_{\text{vir}}$ denotes the total number of
orbitals under consideration, the generalized eigenvalue problem of
pp-RPA is given by 
\begin{equation}\label{eq:ppRPA}
  \begin{pmatrix}
    A & B \\
    B^{\trans} & C 
  \end{pmatrix}
  \begin{pmatrix} 
    X \\
    Y 
  \end{pmatrix} 
  = 
  \omega
  \begin{pmatrix}
    I_p \\
    & - I_h
  \end{pmatrix}
    \begin{pmatrix} 
    X \\
    Y 
  \end{pmatrix}, 
\end{equation}
where $I_p$ and $I_h$ are identity matrices of dimension
$N_p=\binom{N_{\text{occ}}}{2}$ and $N_h=\binom{N_{\text{vir}}}{2}$,
respectively, and entries in matrices $A$, $B$, and $C$ are defined
via
\begin{align*}
  & A_{ijkl} = \bra{ij} \ket{kl} + \delta_{ik}\delta_{jl} (\eps_i + \eps_j - 2 \eps_F), \\
  & B_{ijcd} = \bra{ij} \ket{cd}, \\
  & C_{abcd} = \bra{ab} \ket{cd} - \delta_{ac}\delta_{bd} (\eps_a +
  \eps_b - 2 \eps_F),
\end{align*}
for $1\leq j<i\leq N_{\text{occ}}$, $1\leq l<k\leq N_{\text{occ}}$, $N_{\text{occ}}+1\leq b<a\leq N$, and $N_{\text{occ}}+1\leq d<c\leq N$, where 
\begin{align*}
  & \bra{pq} \ket{rs} = \braket{pq}{rs} - \braket{pq}{sr}, 
\end{align*}
and 
\begin{align*}
  & \braket{pq}{rs} = \iint_{\TT \times \TT} \phi_p(r_1) \phi_q(r_2) \phi_r(r_1) \phi_s(r_2) v_c(r_1 - r_2)  \ud r_1 \ud r_2
\end{align*}
is the four-center two-electron repulsion integral. Here $v_c(\cdot)$ is the periodic Coulomb kernel (due to our choice of the periodic boundary condition) given by the fundamental solution of the Poisson equation with a periodic boundary condition on $\TT$:
\begin{equation}
  - \Delta v_c(\cdot) = 4 \pi ( \delta(\cdot) - 1 ), 
\end{equation}
where $\delta(\cdot)$ is the Dirac delta function. 

The dimension of pp-RPA matrix 
\begin{equation}
\label{eq:mat}
  \begin{pmatrix}
    A & B \\
    B^{\trans} & C 
  \end{pmatrix}
\end{equation}
is $O(N^2) \times O(N^2)$; and thus constructing the whole pp-RPA
matrix takes at least $O(N^4)$ operations, since it contains $O(N^4)$
entries. The action of this matrix to a vector also scales as $O(N^4)$
in general. Thus, the standard approach for the generalized eigenvalue
problem \eqref{eq:ppRPA} has a computational cost at least $O(N^4)$
for getting a single eigenpair. 

In this work, we propose an $O(N^3)$ scaling algorithm to obtain a few
eigenpairs of the generalized eigenvalue problem above. The observation is
that if an iterative algorithm such as the Jacobi-Davidson eigensolver
\cites{JDgeig,Gerard2000} is used, the computational bottleneck is to
apply the pp-RPA matrix to a vector (referred as \textsf{matvec} in
the sequel); in particular, it is not necessary to construct the
matrix for \textsf{matvec}. An $O(N^3)$ \textsf{matvec} is available
by an efficient representation of the electron repulsion integral
tensor enabled by the recently proposed interpolative separable
density fitting in \cites{Lu2015, Lu2015Preprint}. Combined with the
Jacobi-Davidson iterative eigensolver, this gives a cubic scaling
algorithm for the pp-RPA excitation energy calculation.

The rest of the paper is organized as follows. Section \ref{sub:df}
introduces an $O(N^2N_{\text{grid}})$ interpolative separable density
fitting (ISDF).  Section \ref{sub:matvec} describes an
$O(N N_{\text{aux}}^2+N^2N_{\text{aux}})$ \textsf{matvec} based on the
results of the ISDF. Section \ref{sub:JD} briefly revisits the
Jacobi-Davidson eigensolver and discusses a preconditioner for
applying to pp-RPA. Section \ref{sub:app} proposes a truncated pp-RPA
model to reduce the prefactor of our cubic scaling
algorithm. Numerical examples are provided in Section
\ref{sec:numerics} to support the proposed algorithm.


\section{Algorithm}
\label{sec:algo}

In this section, we describe the proposed cubic scaling algorithm in
detail. 
In what follows, we
will use capital letters to denote matrices, e.g., $A(x,y)$ represents
a matrix denoted by $A$ with row index $x$ and column index $y$,
$A^{\trans}$ is the transpose of $A$, and $A^*$ is the complex
conjugate transpose of $A$.

\subsection{Interpolative separable density fitting}
\label{sub:df}

Recall that the pp-RPA matrix \eqref{eq:mat} involves the four-center two-electron repulsion integrals for a given set of orbitals $\{\phi_p\}_{1\leq p\leq N}\subset L^2(\TT)$ as 
\begin{equation*}
  \braket{pq}{rs} = \iint_{\TT \times \TT} \phi_p(x) \phi_q(y) \phi_r(x) \phi_s(y) v_c(x - y) \ud x \ud y.
\end{equation*}
To obtain such integrals for all possible $p, q, r, s$, we can first evaluate 
\footnote{Throughout this work, we write $qs$ as a pair index, instead of the product of $q$ and $s$.}
\begin{equation}\label{eq:diag1}
  V_{qs}(x) = \int_{\TT} \phi_q(y) \phi_s(y) v_c(x - y) \ud y
\end{equation}
using FFT with cost $O(N^2 N_{\grid})$. The repulsion integral can then be obtained as 
\begin{equation}\label{eq:diag2}
  \braket{pq}{rs} = \int_{\TT} \phi_p(x) \phi_r(x) V_{qs}(x) \ud x, 
\end{equation} 
which scales as $O(N^4 N_{\grid})$.  The $O(N^4N_{\text{grid}})$
scaling makes it prohibitively expensive to construct the pp-RPA
matrix if $N$ (and hence $N_{\grid}$) is large, which motivates the
development of efficient representation of the electron repulsion
integral, in particular the density fitting approach (also known as
the resolution of identity approach) for pair density (see e.g.,
\cite{Dunlap1979,Xinguo2012,Sodt2006,Weigend1998143}).

The idea behind the density fitting approach is to explore the
(numerically) low-rank structure of the pair density, viewed as a
matrix with indices $(pq, x)$:\footnote{With some abuse of notation,
  in this paper we do not distinguish the spatial variable $x$ with
  the index of spatial grid.}
\begin{equation*}
\Phi_{pq}(x)=\phi_p(x)\phi_q(x)\in \mathbb{R}^{N^2\times
  N_{\text{grid}}}.
\end{equation*}
Viewing the periodic Coulomb kernel as a matrix $v_c(x, y)$, the
electron repulsion integrals can be considered as entries in the matrix $\Phi v_c \Phi^\trans$. Therefore, if we could find a low-rank approximation of $\Phi$ in the sense that
\begin{equation}\label{eq:lr}
\Phi_{pq}(x) \approx \sum_\mu S_{pq}^\mu P_\mu(x),
\end{equation}
where $\mu=1,2,\dots,N_{\text{aux}}$ labels the auxiliary basis
functions $\{P_\mu(x)\}$, with $N_{\aux} = O(N)$, then the electron
repulsion integrals can be represented as
\begin{equation}\label{eq:nsp1}
  \braket{pq}{rs} \approx \sum_{\mu\nu} V(\mu,\nu) S_{pr}^\mu S_{qs}^\nu,
\end{equation}
and
\begin{equation}\label{eq:nsp2}
  \braket{pq}{sr} \approx \sum_{\mu\nu} V(\mu,\nu) S_{ps}^\mu S_{qr}^\nu,
\end{equation}
where $V(\mu,\nu) =\sum_{x,y}P_\mu(x)v_c(x,y)P_\nu(y)$. The drawback of the density fitting approach though is that the factor $S$ introduced in \eqref{eq:lr} 
remains to be a large $N^2 \times N_{\aux}$ matrix. 
This leads to higher computational complexity when
applying the pp-RPA matrix to a vector. This can be understood as the
indices $pqrs$ in the representation \eqref{eq:nsp1} and
\eqref{eq:nsp2} are not separable.

The interpolative separable density fitting (ISDF) was proposed in
\cite{Lu2015,Lu2015Preprint} aiming at a more efficient
representation. The main idea is to apply a randomized column
selection algorithm \cite{Liberty18122007} to obtain a low-rank
interpolative decomposition such that columns in $S$ are actually 
the important columns of $\Phi$, i.e., we obtain a subset $\{\mu\}$
in the spacial grid points $\{x\}$ such that we have a rank-one factorization
\[
S_{pq}^\mu =  \phi_p(\mu)\phi_q(\mu)
\]
for a fixed $\mu$, and a low-rank approximation 
\begin{equation*} 
  \Phi_{pq}(x) \approx \sum_{\mu} \phi_p(\mu)\phi_q(\mu) P_{\mu}(x),
\end{equation*}
where the number of grid points of the subset $\{\mu\}$ is
$N_{\aux} = O(N)$. Denote
$M\in \mathbb{R}^{N\times N_{\aux}}$ the matrix consisting of
$\{\phi_p(\mu)\}_{1\leq p\leq N}$ as its rows, we have
\begin{equation*}
\Phi_{pq}(x)\approx \sum_{\mu}M_p(\mu)M_q(\mu)P_\mu(x).
\end{equation*}
Hence, once the interpolative separable density fitting is available, the repulsion integrals can be represented via the tensor hypercontraction format  \cite{Hohenstein,Parrish2012}
\begin{equation*}
  \braket{pq}{rs} \approx \sum_{\mu\nu} V(\mu,\nu) M_p(\mu)  M_r(\mu)  M_q(\nu) M_s(\nu),
\end{equation*}
and similarly
\begin{equation*}
  \braket{pq}{sr} \approx \sum_{\mu\nu} V(\mu,\nu) M_p(\mu) M_s(\mu)M_q(\nu) M_r(\nu) .
\end{equation*}
The main difference between the above two equations and those in
\eqref{eq:nsp1}--\eqref{eq:nsp2} is the separable dependence on the
indices $p$, $q$, $r$, and $s$. As we shall see later, taking
advantage of this separable dependence is the key idea for a fast
\textsf{matvec} to apply the pp-RPA matrix.

Direct construction of an ISDF of $\Phi$ is expensive since $\Phi$ is
a large matrix of size $N^2\times N_{\text{grid}}$. Instead, the
method in \cite{Lu2015Preprint} chooses $O(\sqrt{N})$ representative
row vectors from the resulting matrix of a random linear combination of 
\begin{equation}\label{eq:phibefore}
[\phi_1(x), \phi_2(x), \dots, \phi_N(x)]^T \in \mathbb{R}^{N\times
  N_{\text{grid}}}.
  \end{equation}
These $O(\sqrt{N})$ representative row vectors form a matrix $U$ of size
$O(\sqrt{N}\times N_{\text{grid}})$ as a compressed representation of the matrix in \eqref{eq:phibefore}. 
Instead of working on the ISDF of of $\Phi$ of size $N^2\times N_{\text{grid}}$, it is cheaper to 
construct the ISDF of the matrix
\[
\Xi(ij,x)=\wb{U}(i,x)U(j,x)
\]
of size $O(N)\times N_{\text{grid}}$. 

Detailed algorithms in \cite{Lu2015Preprint} are recalled
below.  An auxiliary column selection algorithm is given in Algorithm
\ref{alg:qr} and the main algorithm for interpolative separable
density fitting is described in Algorithm \ref{alg:df}. In these
algorithms, we will adopt \textsf{MATLAB} notations for submatrices.

\begin{algorithm2e}[H]
\label{alg:qr}
\caption{Column selection algorithm based on pivoted QR.}
\SetKwInOut{Input}{Input}
\SetKwInOut{Output}{Output}
\Input{A matrix $M\in \mathbb{C}^{ m\times n}$, error tolerance $\epsilon$}
\Output{An $m\times n_0$ submatrix $M_0$ of $M$ and an $n_0\times n$ matrix $P$, such that $M\approx M_0P$}
    
Compute the pivoted QR decomposition $[Q,R,E]=\mathsf{qr}(M)$, i.e., 
\[
QR=ME,
\]
where $E$ is an $n\times n$ permutation matrix, $Q$ is an $m\times m$ unitary matrix, and $R$ is an $m\times n$ upper triangular matrix with diagonal entries in decreasing order;

Set $n_0$ such that
\[
|R(n_0,n_0)|\geq \epsilon \, |R(1,1)|>|R(n_0+1,n_0+1)|;
\]

Set $M_0=(ME)(:,1:n_0)$, the first $n_0$ columns of $ME$;

Compute $P=R^{-1}(1:n_0,1:n_0)R(1:n_0,:)E^{-1}$.

\end{algorithm2e}

\begin{algorithm2e}
\label{alg:df}
\caption{Interpolative separable density fitting.}
\SetKwInOut{Input}{Input}
    \SetKwInOut{Output}{Output}
    \Input{Orbitals $\{\phi_p(x)\}_{1\leq p\leq N}$, error tolerance $\epsilon$, and a column selection parameter $c$}
    \Output{Selected grid points $\{\mu\}\subset \{x\}$ and an auxiliary matrix $S$, such that
    \[
\Phi(pq,x)=\phi_p(x)\phi_q(x) \approx \sum_{\mu}M_p(\mu) M_q(\mu)S(\mu,x).
\]}

Reshape the orbital functions $\phi_p(x)$ as a matrix $\phi(p, x)$;

Compute the discrete Fourier transform of $\phi$ left multiplied by a random diagonal matrix:
\[
\widehat{\phi}(\xi,x)=\sum_{p=1}^{N} e^{-2\pi i \xi p/(N)}\eta_p \phi(p,x),
\]
for all $\xi$, $1\leq \xi\leq N$, where $\eta_p$ is a random unit complex number for each $p$;

Choose a submatrix $U$ of $\widehat{\phi}$ by randomly choosing $r=c\sqrt{N}$ rows;

Construct an $r^2\times N$ matrix $\Xi$ such that 
\[
\Xi(ij,x)=\wb{U}(i,x)U(j,x)
\]
for all $x$, $1\leq i\leq r$, and $1\leq j\leq r$, where $(ij)$ is viewed as the row index of $M$ instead of the product of $i$ and $j$;

Apply Algorithm \ref{alg:qr} on the $r^2\times N$ matrix $\Xi$ with the parameter $\epsilon$ to find important columns of $\Xi$ with indices $\{\mu\}\subset\{x\}$ and an auxiliary matrix $S$, such that
\[
\Xi(pq,x)\approx \sum_{\mu}\Xi(pq,\mu)S(\mu,x);
\] 

Find the submatrix $M$ of $D$ with column indices $\{\mu\}$ and finally we have 
    \[
\Phi(pq,x)=\phi_p(x)\phi_q(x) \approx \sum_{\mu}M_p(\mu) M_q(\mu)S(\mu,x).
\]
\end{algorithm2e}

The computational cost in Algorithm \ref{alg:qr} is $O(m^2 n)$,
dominated by the QR factorization of $M$, since $n_0$ (the number of
columns selected) is assumed to be smaller than $m$ or $n$ and we have
assumed $m \leq n$. In Algorithm \ref{alg:df}, the dominant cost is
the application of Algorithm \ref{alg:qr} on a matrix of size
$O(N) \times N_{\text{grid}}$ in Step $5$, since other steps take
operations less than $N^2 N_{\text{grid}}$. In sum, given a set of
orbitals $\{\phi_p(x)\}_{1\leq p\leq N}$, the computational cost to
obtain the interpolative separable density fitting is
$O(N^2N_{\text{grid}})$ operations. 

\subsection{Cubic scaling \textsf{matvec}}
\label{sub:matvec}

In the previous subsection, we have introduced Algorithm \ref{alg:qr}
and \ref{alg:df} to construct the interpolative separable density
fitting from a set of given orbitals $\{\phi_p(x)\}_{1\leq p\leq N}$.
The output of Algorithm \ref{alg:df} is a set of selected grid points
$\{\mu\}\subset \{x\}$, compressed orbitals $\{M_p(\mu)\}$, and an
auxiliary matrix $P$, such that
\[
\Phi(pq,x)=\phi_p(x)\phi_q(x) \approx \sum_{\mu}M_p(\mu) M_q(\mu)P_\mu(x).
\]
An immediate result of the above equation is the following interpolative separable density fitting for repulsion integrals
\begin{equation}\label{eq:decompRI1}
  \braket{pq}{rs} \approx \sum_{\mu\nu} V(\mu,\nu) M_p(\mu) M_r(\mu)  M_q(\nu) M_s(\nu) ,
\end{equation}
and similarly
\begin{equation}\label{eq:decompRI2}
  \braket{pq}{sr} \approx \sum_{\mu\nu} V(\mu,\nu) M_p(\mu) M_s(\mu)M_q(\nu) M_r(\nu) .
\end{equation}
We now exploit this representation for a cubic scaling \textsf{matvec} of the pp-RPA matrix. 

When we apply the pp-RPA matrix
\begin{equation}
\label{eq:dv}
  \begin{pmatrix}
    A & B \\
    B^{\trans} & C 
  \end{pmatrix}
  = 
  \begin{pmatrix} 
    D_p \\
    & D_h 
  \end{pmatrix}
  + 
  \begin{pmatrix} 
  \bra{ij} \ket{kl} & \bra{ij} \ket{cd} \\
    \bra{ab} \ket{kl} &  \bra{ab} \ket{cd} 
  \end{pmatrix}
\end{equation}
to a vector $(g, h)^{\trans}$ where
$g\in \mathbb{R}^{N_{\text{occ}}(N_{\text{occ}}-1)/2}$ and
$h\in \mathbb{R}^{N_{\text{vir}}(N_{\text{vir}}-1)/2}$, the action of
the diagonal matrices $D_p$ and $D_h$ is simple. Thus, let us focus on
how to compute $(A-D_p)g$, $Bh$, $B^{\trans}g$, and $(C-D_h)h$. Since
the entries in $A$, $B$, and $C$ have similar definitions and
structures, it is sufficient to illustrate how to compute $(A-D_p)g$
with cubic scaling.

By definition
\begin{equation*}
 (A-D_p)_{ijkl} = \bra{ij} \ket{kl} 
\end{equation*}
with $(ij)$ as the row index of $A-D_p$ and $(kl)$ as its column index
for $1\leq j<i\leq N_{\text{occ}}$ and $1\leq l<k\leq N_{\text{occ}}$.
Hence, we also use $(kl)$ as the row indices of
$g\in\mathbb{R}^{N_{\text{occ}}(N_{\text{occ}}-1)/2}$. Taking the
representation of the electron repulsion integral
\eqref{eq:decompRI1}--\eqref{eq:decompRI2}, we have
\begin{equation}\label{eq:app1}
  \begin{aligned}
    \sum_{kl} \braket{ij}{kl} g_{kl} & = 
    \sum_{kl} \sum_{\mu\nu} V(\mu,\nu) M_i(\mu) M_k(\mu) M_j(\nu) M_l(\nu)  g_{kl} \\
    & = \sum_{\mu} M_i(\mu) \Biggl( \sum_{\nu} M_j(\nu) V(\mu,\nu)  \sum_k M_k(\mu) \biggl( \sum_l M_l(\nu) g_{kl} \biggr) \Biggr), 
  \end{aligned}
\end{equation}
and similarly 
\begin{equation}\label{eq:app2}
  \begin{aligned}
    \sum_{kl} \braket{ij}{lk} g_{kl} & = 
    \sum_{kl} \sum_{\mu\nu} V(\mu,\nu) M_i(\mu) M_l(\mu) M_j(\nu) M_k(\nu)  g_{kl} \\
    & = \sum_{\mu} M_i(\mu) \Biggl( \sum_{\nu} M_j(\nu) V(\mu,\nu)  \sum_l M_l(\mu) \biggl( \sum_k M_k(\nu) g_{kl} \biggr) \Biggr). 
  \end{aligned}
\end{equation}

The key observation is that all the above calculation for all index
pairs $(ij)$, $1\leq j<i\leq N_{\text{occ}}$, can be done in cubic
scaling cost as follows. Let us take the summation
$ \sum_{kl} \braket{ij}{kl} g_{kl} $ as an example, the algorithm goes
as follows
\begin{itemize}
\item Step $1$: compute $E(k,\nu) =\sum_l M_l(\nu) g_{kl}$ for all $\nu$ and $k$ with $O(N^2N_{\aux})$ operations;
\item Step $2$: compute $F(\mu,\nu) = \sum_k M_k(\mu) E(k,v)$ for all $\mu$ and $\nu$ with $O(NN_{\aux}^2)$ operations;
\item Step $3$: compute $G(j,\mu)= \sum_{\nu} M_j(\nu) V(\mu,\nu) F(\mu,\nu)$ for all $\mu$ and $j$ with $O(NN_{\aux}^2)$ operations;
\item Step $4$: compute
  $ \sum_{kl} \braket{ij}{kl} g_{kl}= \sum_{\mu} M_i(\mu) G(b,\mu)$
  for all $i$ and $j$ with $O(N^2N_{\aux})$ operations.
\end{itemize}
Similarly, $\sum_{kl} \braket{ij}{lk} g_{kl} $ for all index pairs $(ij)$ can be computed in $O(NN_{\aux}^2+N^2N_{\aux})$ operations too. 

In sum, we have obtained an $O(NN_{\aux}^2+N^2N_{\aux})$ algorithm to evaluate
$(A-D_p)g$. As a result, we have a cubic scaling \textsf{matvec} for
$Ag$.  We can compute $Bh$, $B^{\trans}g$, and $Ch$ similarly and
arrive at an $O(NN_{\aux}^2+N^2N_{\aux})$ \textsf{matvec} to apply the pp-RPA matrix.

\subsection{Jacobi-Davidson eigensolver}
\label{sub:JD}


The Jacobi-Davidson generalized eigensolver \cite{JDgeig,Gerard2000}
is a matrix-free method (e.g., only \textsf{matvec} operation is
required) and allows to use a preconditioner for solving linear
systems in its inner iteration to accelerate the overall
convergence. For completeness, a detailed description of the algorithm
is given in Appendix~\ref{app}.

Empirically we observe in our numerical examples that pp-RPA
matrices are usually strongly diagonally dominant (although this
observation has not been verified in theory), the diagonal part of
these matrices can be chosen as the preconditioner of the
Jacobi-Davidson eigensolver. Note that the computational cost for all
repulsion integrals in the diagonal part takes
$O(N^2 N_{\text{grid}})$ operations and memory by \eqref{eq:diag1} and
\eqref{eq:diag2}. To avoid this expensive computation and memory
request, we choose the following diagonal matrix as the preconditioner
instead of the exact diagonal of the pp-RPA matrix:
\begin{equation}\label{eq:precond}
P =   \begin{pmatrix} 
    D_p \\
    & D_h 
  \end{pmatrix},
\end{equation}
where $D_p$ and $D_h$ are defined in \eqref{eq:dv}. 

In each iteration of the Jacobi-Davidson eigensolver, the dominant cost 
is applying the pp-RPA matrix to $O(1)$ vectors and solving 
one linear system of a shifted pp-PRA matrix. Since the GMRES \cite{GMRES} 
is applied to solve this linear system with a fixed number of iterations, 
the complexity of the linear system solver is the same as that of the 
\textsf{matvec}, which is $O(N^2 N_{\text{aux}}+N_{\text{aux}}^2N)$. 
Since the number of iterations in the eigensolver depends on the 
accuracy of GMRES and we have fixed the number of iterations in GMRES, 
a good preconditioner is important to improve the convergence 
of the eigensolver. As we shall see later 
in Section \ref{sec:numerics}, numerical examples show that the 
preconditioner in \eqref{eq:ppRPA} is sufficiently good to reduce the number of iterations 
in the Jacobi-Davidson eigensolver and keep this number roughly 
independent of the problem size. Therefore, our proposed algorithm can 
compute $O(1)$ eigenpairs with $O(N^2 N_{\text{aux}}+N_{\text{aux}}^2N)$ operations.


\subsection{Truncation of orbital space in the pp-PRA model}
\label{sub:app}
In the original pp-RPA model proposed in \cite{Aggelen2013,Peng2013,Scuseria2013}, the pp-RPA matrix involves all orbitals $\{\phi_p(x)\}_{1\leq p\leq N_{\text{grid}}}$ of the $N_{\text{grid}}\times N_{\text{grid}}$ Hamiltonian matrix $H$. Recall its definition
\begin{equation}\label{eq:appRPA}
  \begin{pmatrix}
    A & B \\
    B^{\trans} & C 
  \end{pmatrix}
  \begin{pmatrix} 
    X \\
    Y 
  \end{pmatrix} 
  = 
  \omega
  \begin{pmatrix}
    I_p \\
    & - I_h
  \end{pmatrix}
    \begin{pmatrix} 
    X \\
    Y 
  \end{pmatrix}, 
\end{equation}
where $I_p$ and $I_h$ are identity matrices of size $p=\binom{N_{\text{occ}}}{2}$ and $h=\binom{N_{\text{vir}}}{2}$ (where $N_{\text{vir}} + N_{\text{occ}}=N= N_{\text{grid}}$), respectively, and entries in matrices $A$, $B$, and $C$ are defined via
\begin{align*}
  & A_{ijkl} = \bra{ij} \ket{kl} + \delta_{ik}\delta_{jl} (\eps_i + \eps_j - 2 \eps_F), \\
  & B_{ijcd} = \bra{ij} \ket{cd}, \\
  & C_{abcd} = \bra{ab} \ket{cd} - \delta_{ac}\delta_{bd} (\eps_a +
  \eps_b - 2 \eps_F),
\end{align*}
for all $1\leq j<i\leq N_{\text{occ}}$, $1\leq l<k\leq N_{\text{occ}}$, $N_{\text{occ}}+1\leq b<a\leq N_{\text{grid}}$, and $N_{\text{occ}}+1\leq d<c\leq N_{\text{grid}}$. The lowest (the smallest magnitude) eigenpairs of the above generalized eigenvalue problem is able to  predict electronic excited states. Note that
\begin{enumerate}
\item the pp-RPA matrix is usually strongly diagonally dominant; and 
\item the diagonal entries of the pp-RPA matrix is essentially governed by $(\eps_i + \eps_j - 2 \eps_F)$ and $ (\eps_a +  \eps_b - 2 \eps_F)$.
\end{enumerate}
It is reasonable to reduce the system size of the pp-RPA equation by
only keeping a small portion of orbitals with orbital energy close to
the Fermi energy $\eps_F$, while maintaining the lowest eigenvalues
approximately the same. In other words, we can use ${N}_{\text{vir}}$
virtual orbitals with orbital energy closest to $\eps_F$ and
${N}_{\text{occ}}$ occupied orbitals with orbital energy closest to
$\eps_F$ in the construction of the pp-RPA matrix. We will test
${N}_{\text{occ}}$ far less than the total number of occupied orbitals
and ${N}_{\text{vir}}$ far less than the total number of virtual
orbitals, i.e.,
$N={N}_{\text{occ}}+{N}_{\text{vir}}\ll {N}_{\text{grid}}$. This
reduced system leads to a much smaller prefactor of the cubic scaling
algorithm while keeping the accuracy of the excited state
prediction. Numerical examples with varying problem sizes in Section
\ref{sec:numerics} show that it is sufficient to use $10$ percents of
the occupied and virtual orbitals to keep four-digit relative accuracy
in estimating the first three positive eigenvalues close to zero and
the first three negative eigenvalues close to zero. Note that a
different active-space truncation of the pp-RPA matrix was proposed
very recently in \cite{ZhangYang}, which can be combined with the truncation
studied here and will be explored in future works.

\section{Numerical examples}\label{sec:numerics}

We now present numerical results to support the efficiency of the
proposed algorithm. In the first part, we verify the cubic scaling of
the \textsf{matvec} proposed in Section \ref{sub:matvec}. In the
second part, we show that the proposed preconditioner in
\eqref{eq:precond} is able to keep the number of iterations in the
Jacobi-Davidson eigensolver roughly independent of the problem
size. Finally, we provide various examples to support the truncation
of orbital space in pp-RPA as in Section \ref{sub:app}.

\subsection{Tests for the cubic scaling \textsf{matvec}}
\begin{figure}
\begin{center}
   \begin{tabular}{c}
        \includegraphics[height=2in]{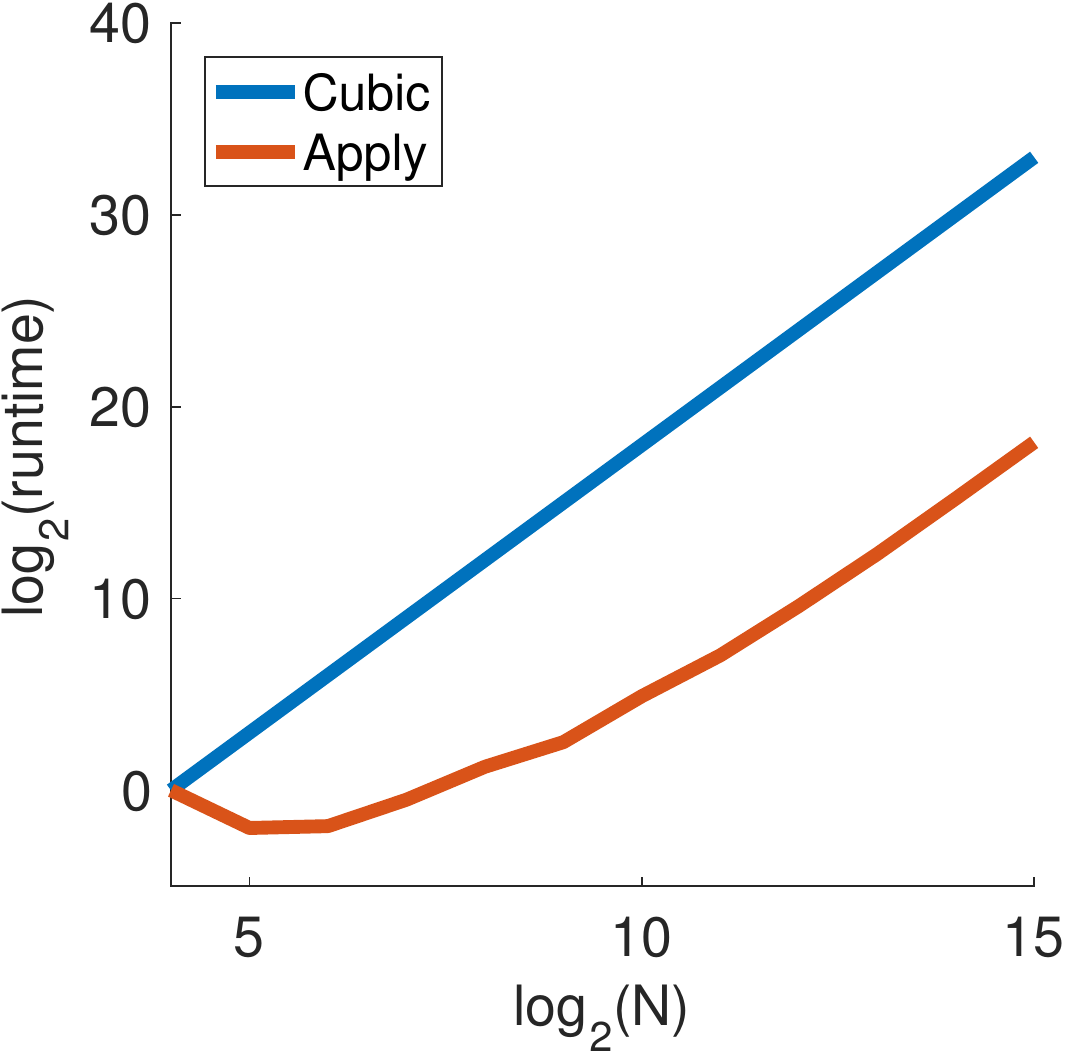}  
  \end{tabular}
\end{center}
\caption{The scaling test of the evaluation of Equation \eqref{eq:app1} with an $N\times N$ matrix $M$, an $N\times N$ matrix $V$, and an $N^2\times 1$ vector $g_{kl}$. The runtime of the evaluation is recorded for different problem sizes $N$. Red: $\log_2(\text{runtime})$ as a function in $\log_2(N)$. Blue: ground truth cubic scaling line as a reference. The scaling is cubic when $N$ is sufficiently large.}
\label{fig:scaling}
\end{figure}

In the first part, we verify that the \textsf{matvec} proposed in Section \ref{sub:matvec} is cubic scaling numerically. By the interpolative separable density fitting technique in \cite{Lu2015Preprint}, we construct small matrix factors $M$, select a set of important spacial grid points $\{\mu\}\subset\{x\}$, and compute the \textsf{matvec} of the pp-RPA matrix via the method detailed in Equation \eqref{eq:app1} and \eqref{eq:app2}. It has been shown in \cite{Lu2015Preprint} that the construction of $M$ and the selection of $\{\mu\}$ take $O(N^2 N_{\text{grid}})$ operations. Hence, to verify the cubic scaling of the \textsf{matvec}, it is sufficient to show that for an $N\times N$ matrix $M$, an $N\times N$ matrix $V$, and an $N^2\times 1$ vector $g_{kl}$,  the computation of Equation \eqref{eq:app1} takes $O(N^3)$ operations for $N^2$ index pairs $(ij)$. Therefore, we generate $M$, $V$, and $g_{kl}$ randomly with different values of $N$, evaluate Equation \eqref{eq:app1}, and summarize the runtime in Figure \ref{fig:scaling}. As shown in Figure \ref{fig:scaling}, the evaluation of Equation \eqref{eq:app1} is cubic scaling as soon as $N$ is sufficiently large. Therefore, the cubic scaling \textsf{matvec} for apply the pp-RPA matrix has been verified.

\subsection{Tests for the preconditioner}\label{sub:prec}
\begin{figure}
\begin{center}
   \begin{tabular}{cc}
        \includegraphics[height=2in]{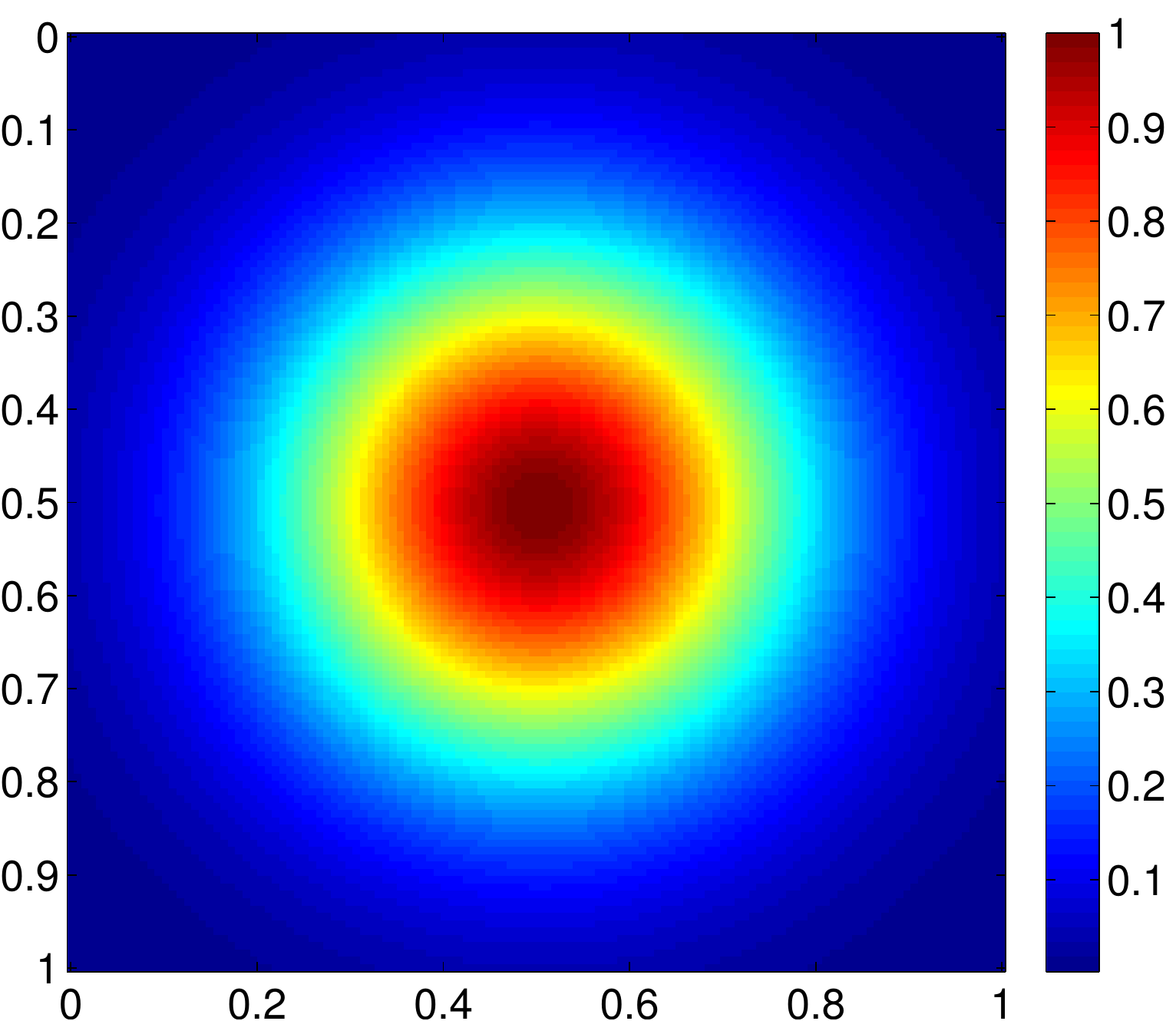}&   \includegraphics[height=2in]{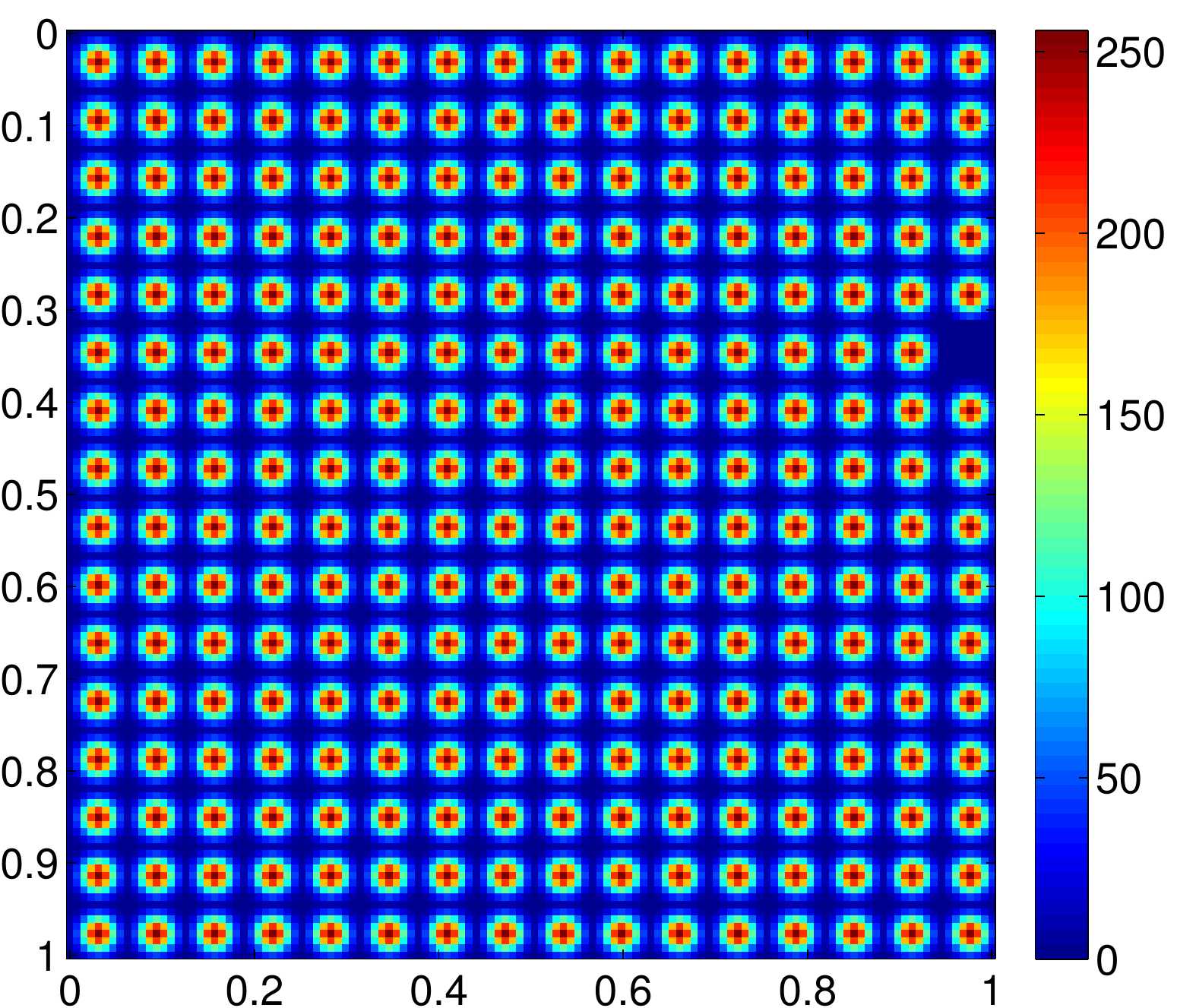} 
  \end{tabular}
\end{center}
\caption{Left: the periodic function $V_0(\mathbf{r})$ is a Gaussian well on the unit square $[0,1)^2$. Right: the potential energy operator $\ell^2 V(\mathbf{r})$ with $\ell=16$.}
\label{fig:pot1}
\end{figure}

In the second part, we use synthetic Hamiltonian to construct approximate pp-RPA matrices. The Hamiltonian matrix $H$ is a discrete representation of the Hamiltonian operator
\begin{equation}
\label{eqn:ex0}
\left( -\frac{\Delta}{2} + V(\mathbf{r})\right) \phi_j(\mathbf{r})=\epsilon_j\phi_j(\mathbf{r}),\qquad \mathbf{r}\in \ell \mathbb{T}^d :=[0,\ell)^d,
\end{equation}
with a periodic boundary condition and $d=1$ or $2$, where $V(\mathbf{r})$ is the potential field, $\epsilon_j$ is the orbital energy of the corresponding Kohn-Sham orbital, $\phi_j(\mathbf{r})$. It is convenient to rescale the system to the unit square via the transformation $ \ell \mathbf{x} =\mathbf{r}$: 
\begin{equation}
\label{eqn:ex1}
\left( -\frac{\Delta}{2} + \ell^d V(\ell\mathbf{x})\right) \phi_j(\mathbf{x})=\epsilon_j \ell^d \phi_j(\mathbf{x}),\qquad \mathbf{x}\in \mathbb{T}^d :=[0,1)^d,
\end{equation}
and discretize the new system with the pseudo-spectral method. Let $V_0(\mathbf{r})$ be a Gaussian well on the unit domain $[0,1)^d$ (see Figure \ref{fig:pot1} (left) for an example when $d=2$) and extend it periodically with period $1$ to obtain $V(x)$ defined on $\ell \mathbb{T}^d :=[0,\ell)^d$. We randomly remove one Gaussian well from $V(x)$ to construct a non-trivial potential field and rescale it to $ \ell^d V(\ell\mathbf{x})$ on the unit domain $[0,1)^d$ (see Figure \ref{fig:pot1} (right) for a two-dimensional example). The number of grid points per dimension within one period is set to $4$. Once the Hamiltonian is available, we compute its eigenpairs by direct diagonalization to obtain its orbitals
$\{\phi_p\}_{1\leq p\leq N_{\text{grid}}}$ and the corresponding orbital energy $\{\epsilon_p\}_{1\leq p\leq N_{\text{grid}}}$. 

We will use one-dimensional Hamiltonian matrices (i.e. $d=1$) to verify the efficiency of the proposed preconditioner in \eqref{eq:precond}. The Jacobi-Davidson eigensolver in Algorithm \ref{alg:eig} in the appendix is applied to compute the generalized eigenvalue closest to zero in the generalized eigenvalue problem  \eqref{eq:ppRPA}, without any preconditioner and with the preconditioner in \eqref{eq:precond}. Parameters in Algorithm \ref{alg:eig} are $\epsilon=1e-10$, $m_{\min}=k_{\max}+5$, $m_{\max}=m_{\min}+5$, and $\text{mx}=400 k_{\max}$. The initial nontrivial vector $\nu_0$ is one realization of a random vector such that each entry is a random variable with a uniform distribution in $[0,2]$.

Table \ref{tab:0} and \ref{tab:00} summarize the number of iterations and the accuracy of the eigensolver, respectively. As in the generalized eigenvalue problem \eqref{eq:ppRPA}, suppose the generalized eigenpair computed is $\left(\omega, \bigl(\begin{smallmatrix} 
    X \\
    Y 
  \end{smallmatrix}\bigr) \right)$, the accuracy in Table \ref{tab:00} is defined to be the $2$-norm  
\begin{equation*}
  \Norm{
  \begin{pmatrix}
    A & B \\
    B^{\trans} & C 
  \end{pmatrix}
  \begin{pmatrix} 
    X \\
    Y 
  \end{pmatrix} 
  -
  \omega
  \begin{pmatrix}
    I_p \\
    & - I_h
  \end{pmatrix}
    \begin{pmatrix} 
    X \\
    Y 
  \end{pmatrix}}.
\end{equation*}
These results show that the proposed preconditioner is able to keep the number of iterations in the eigensolver roughly independent of the problem size, and the accuracy is usually higher in the presence of the preconditioner.

\begin{table}[htp]
\centering
\begin{tabular}{rcccccc}
  \toprule
  $\ell$ & 4 & 8 & 16 & 32 & 64 & 128 \\
\toprule
preconditioned      &    46 &   60 &   56 &   60 &   54 &   57  \\ 
non-preconditioned   & 160 & 246 & 316 & 414 & 581 & 826 \\ 
\bottomrule
\end{tabular}
\medskip
\caption{The number of iterations in the Jacobi-Davidson eigensolver with and without a preconditioner. This table summarizes results for one-dimensional Hamiltonian matrices with different number of Gaussian wells $\ell$. The preconditioned eigensolver needs a number of iterations roughly independent of the problem size.}
\label{tab:0}
\end{table}

\begin{table}[htp]
\centering
\begin{tabular}{rcccccc}
  \toprule
  $\ell$ & 4 & 8 & 16 & 32 & 64 & 128 \\
\toprule
preconditioned      &    1.0e-09 &   1.1e-09 &   1.0e-09 &  1.2e-09 &   1.6e-08 &   5.6e-08  \\ 
non-preconditioned   & 4.5e-10 & 7.1e-09 & 1.0e-08  & 6.0e-08 & 2.2e-02 & 2.4e-07  \\ 
\bottomrule
\end{tabular}
\medskip
\caption{The accuracy of the Jacobi-Davidson eigensolver with and without a preconditioner. This table summarizes results for one-dimensional Hamiltonian matrices with different number of Gaussian wells $\ell$.}
\label{tab:00}
\end{table}

\subsection{Tests for the truncated pp-RPA model}

In this section, we construct Hamiltonian matrices, orbitals, and orbital energies using the same method in Section \ref{sub:prec}. 
We will conduct two sets of test to verify the truncation of orbital spaces in the pp-RPA model. In these tests, the Jacobi-Davidson eigensolver in Algorithm \ref{alg:eig} in the appendix is applied to compute $k_{\max}$ (sufficiently many) generalized eigenvalues close to zero of pp-RPA matrices such that we able to obtain three positive eigenvalues closest to zero and three negative eigenvalues closest to zero. Other parameters in Algorithm \ref{alg:eig} are $\epsilon=1e-10$, $m_{\min}=k_{\max}+5$, $m_{\max}=m_{\min}+5$, and $\text{mx}=400 k_{\max}$. The initial nontrivial vector $\nu_0$ is one realization of a random vector such that each entry is a random variable with a uniform distribution in $[0,2]$.

In the first test, we verify the proposed truncated pp-RPA model by directly constructing the whole pp-RPA matrix in \eqref{eq:appRPA} (i.e., computing the repulsion integrals by naive summations) with different values of $N_{\text{vir}}$ and $N_{\text{occ}}$. When $N_{\text{vir}}$ and $N_{\text{occ}}$ are the total numbers of virtual and occupied orbitals, we obtain the original pp-RPA matrix in \cite{Aggelen2013,Peng2013,Scuseria2013}. Let $\text{pct}$ denote the percentage of occupied and virtual orbitals (with orbital energy closest to $\eps_F$) we used  in the truncated pp-RPA model. We generate truncated pp-RPA matrices with $\text{pct}=0.05$, $0.1$, $0.2$, $0.3$, and $0.4$, compute three smallest positive eigenvalues and three largest negative eigenvalues by the Jacobi-Davidson method (denoted by $\text{val}_{\text{app}}\in \mathbb{R}^6$), compare these eigenvalues with those from the original pp-RPA matrix (denoted by $\text{val}_{\text{org}}\in \mathbb{R}^6$) via the relative difference below 
\begin{equation}\label{eq:err}
\text{err}=\max\bigl\{ |\text{val}_{\text{app}}(k)-\text{val}_{\text{org}}(k)|/|\text{val}_{\text{org}}(k)|\bigr\}_{1\leq k\leq 6}.
\end{equation}
Note that when $N_{\text{grid}}$ is small, $\text{pct}\cdot N_{\text{grid}}$ might be too small to establish a meaningful pp-RPA equation. Therefore, if $N_{\text{vir}}$ and $N_{\text{occ}}$ are smaller than $4$ due to a small $\text{pct}$, we will update $N_{\text{vir}}$ and $N_{\text{occ}}$ to $4$.

Table \ref{tab:1} and Table \ref{tab:2} summarize the relative difference $\text{err}$ in these comparisons for one-dimensional and two-dimensional Hamiltonian matrices, respectively. These results show that, to estimate three smallest positive eigenvalues and three largest negative eigenvalues of the original pp-RPA matrix (i.e., the pp-RPA matrix constructed with all $N_{\text{grid}}$ orbitals using the naive implementation) within $4$-digits accuracy,  it is sufficient to use only $10$ percents of both the occupied and the virtual orbitals to construct a truncated pp-RPA matrix via the naive implementation.

\begin{table}[htp]
\centering
\begin{tabular}{rccccc}
  \toprule
  $\ell\backslash$ pct & 0.05 & 0.1 & 0.2 & 0.3 & 0.4 \\
\toprule
4   & 9.9e-07 & 9.9e-07 & 9.9e-07 & 9.9e-07 & 1.8e-07 \\ 
8   & 1.9e-07 & 1.9e-07 & 1.6e-07 & 1.4e-07 & 1.4e-07 \\ 
16 & 2.8e-08 & 2.3e-08 & 2.2e-08 & 1.6e-08 & 1.0e-08 \\ 
32 & 2.9e-09 & 2.5e-09 & 1.2e-09 & 3.8e-10 & 1.9e-10 \\
\bottomrule
\end{tabular}
\medskip
\caption{The relative difference $\text{err}$ defined in \eqref{eq:err} of three smallest positive eigenvalues and three largest negative eigenvalues of the original pp-RPA matrix and the truncated pp-RPA matrices constructed with different values of $\text{pct}$. Naive implementation is used in both cases. This table summarizes results for one-dimensional Hamiltonian matrices with different number of Gaussian wells $\ell=4$, $8$, $16$, and $32$.}
\label{tab:1}
\end{table}

\begin{table}[htp]
\centering
\begin{tabular}{rccccc}
  \toprule
  $\ell\backslash$ pct & 0.05 & 0.1 & 0.2 & 0.3 & 0.4 \\
\toprule
2  & 3.2e-04 & 1.8e-04 & 5.7e-05 & 2.3e-05 & 1.5e-05 \\ 
3 & 5.0e-06 & 3.4e-06 & 3.2e-06 & 3.1e-06 & 3.1e-06 \\ 
 \bottomrule
\end{tabular}
\medskip
\caption{The relative difference $\text{err}$ defined in \eqref{eq:err} of three smallest positive eigenvalues and three largest negative eigenvalues of the original pp-RPA matrix and the truncated pp-RPA matrices constructed with different values of $\text{pct}$. Naive implementation is used in both cases.  This table summarizes results for two-dimensional Hamiltonian matrices with different number of Gaussian wells $\ell^2=4$ and $9$.}
\label{tab:2}
\end{table}

In the second test, we verify the truncated pp-RPA model by applying the proposed fast \textsf{matvec} to compute the eigenvalues of the truncated pp-RPA matrices with different values of $N_{\text{vir}}$ and $N_{\text{occ}}$. When $N_{\text{vir}}$ and $N_{\text{occ}}$ are the total numbers of virtual and occupied orbitals, we approximately obtain the original pp-RPA matrix in \cite{Aggelen2013,Peng2013,Scuseria2013} up to some error introduced by the density fitting. In the construction of the interpolative separable density fitting, we set the parameter $\epsilon=1e-7$ and $c=10$.  Again, we generate truncated pp-RPA matrices with $\text{pct}=0.05$, $0.1$, $0.2$, $0.3$, and $0.4$, compute three smallest positive eigenvalues and three largest negative eigenvalues by the Jacobi-Davidson method (denoted by $\text{val}_{\text{app}}\in \mathbb{R}^6$), compare these eigenvalues with those from the original pp-RPA matrix (denoted by $\text{val}_{\text{org}}\in \mathbb{R}^6$) by computing the relative difference as in \eqref{eq:err}. Since in the first test, we have computed the ground truth eigenvalues from the exact pp-RPA matrix by direct evaluation, we reuse these eigenvalues if available, instead of using the eigenvalues of the truncated pp-RPA matrix when $\text{pct}=1$.

Table \ref{tab:3} and Table \ref{tab:4} summarize these comparisons for one-dimensional and two-dimensional Hamiltonian matrices, respectively. These results lead to the same conclusion as in the first test. 

\begin{table}[htp]
\centering
\begin{tabular}{rccccc}
  \toprule
  $\ell\backslash$ pct & 0.05 & 0.1 & 0.2 & 0.3 & 0.4 \\
\toprule
4   & 9.9e-07 & 9.9e-07 & 9.9e-07 & 9.9e-07 & 1.8e-07 \\ 
8   & 1.9e-07 & 1.9e-07 & 1.6e-07 & 1.4e-07 & 1.4e-07 \\ 
16   & 2.8e-08 & 2.3e-08 & 2.2e-08 & 1.6e-08 & 1.0e-08 \\ 
32   & 2.9e-09 & 2.5e-09 & 1.2e-09 & 3.8e-10 & 1.9e-10 \\ 
64   & 3.0e-10 & 1.4e-10 & 4.3e-11 & 2.1e-11 & 9.6e-12 \\ 
128   & 1.6e-11 & 6.5e-12 & 2.4e-12 & 5.0e-08 & 5.7e-13 \\ 
256   & 8.3e-13 & 4.0e-13 & 1.5e-13 & 7.0e-14 & 3.5e-14 \\ 
\bottomrule
\end{tabular}
\medskip
\caption{The relative difference $\text{err}$ defined in \eqref{eq:err} of three smallest positive eigenvalues and three largest negative eigenvalues of the original pp-RPA matrix and the truncated pp-RPA matrices constructed with different values of $\text{pct}$. This table summarizes results for one-dimensional Hamiltonian matrices with different number of Gaussian wells $\ell=4$, $8$, $\dots$, and $256$. The cubic scaling algorithm is used for truncated pp-RPA, for the original pp-RPA, naive implementation is used when $\ell=4$, $\dots$, and $32$, and cubic scaling algorithm is used when 
$\ell=64$, $128$, and $256$ since the naive algorithm is too slow. }
\label{tab:3}
\end{table}

\begin{table}[htp]
\centering
\begin{tabular}{rccccc}
  \toprule
  $\ell\backslash$ pct & 0.05 & 0.1 & 0.2 & 0.3 & 0.4 \\
\toprule
2  & 3.2e-04 & 1.8e-04 & 5.7e-05 & 2.3e-05 & 1.5e-05 \\ 
3 & 5.0e-06 & 3.4e-06 & 3.2e-06 & 2.8e-05 & 3.1e-06 \\ 
4 & 2.8e-05 & 2.1e-05 & 1.8e-05 & 1.7e-05 & 1.3e-05 \\ 
5 & 9.1e-05 & 9.1e-05 & 9.1e-05 & 1.2e-06 & 7.7e-07 \\ 
\bottomrule
\end{tabular}
\medskip
\caption{The relative difference $\text{err}$ defined in \eqref{eq:err} of three smallest positive eigenvalues and three largest negative eigenvalues of the original pp-RPA matrix and the truncated pp-RPA matrices constructed with different values of $\text{pct}$. This table summarizes results for two-dimensional Hamiltonian matrices with different number of Gaussian wells $\ell^2=4$, $9$, $16$, and $25$. 
  The cubic scaling algorithm is used for truncated pp-RPA, for the original pp-RPA, naive implementation is used when $\ell^2=4$ and $9$, and cubic scaling algorithm is used when $\ell^2=16$ and $25$ since the naive algorithm is too slow.
}
\label{tab:4}
\end{table}


\appendix

\section{The Jacobi-Davidson eigensolver}
\label{app}

In Algorithm~\ref{alg:eig} we describe the Jacobi-Davidson algorithm
\cite{JDgeig,Gerard2000} to compute $k_{\max}$ generalized eigenpairs
with generalized eigenvalues closest to a target
$\tau$. 
In particular, if low-lying excitations are desired, we can take
$\tau=0$ in the algorithm.
The algorithm description follows the lecture note by Peter Arbenz available at  \url{http://people.inf.ethz.ch/arbenz/ewp/Lnotes/chapter12.pdf}, and the codes are available at Gerard
L.G.~Sleijpen's personal homepage: \url{http://www.staff.science.uu.nl/~sleij101/}.

\begin{algorithm2e}[ht]
\label{alg:eig}
\caption{Jacobi-Davidson QZ method for $k_{\max}$ interior eigenvalues close to $\tau$.}

\SetKwInOut{Input}{Input}
    \SetKwInOut{Output}{Output}
    \Input{Square matrices $A$ and $B$ in $\mathbb{C}^{n\times n}$, $\tau$, $k_{\max}$, the accuracy parameter $\epsilon$, restart parameters $m_{\min}$ and $m_{\max}$, the maximum number of iterations $\text{mx}$.}
    \Output{$Q$ and $Z\in \mathbb{C}^{n\times k_{\max}}$, $R^A$ and $R^B\in \mathbb{C}^{k_{\max}\times k_{\max}}$ s.t.$AQ=ZR^A$ and $BQ=ZR^B$. The $k_{\max}$ interior generalized eigenvalues close to the target $\tau$ are $\{ R^A_{k,k}/R^B_{k,k} \}_{1\leq k\leq k_{\max}}$.}

Choose a nontrivial vector $v_0$; $k=0$; $\nu_0=1/\sqrt{1+|\tau|^2}$; $\mu_0=-\tau \nu_0$; $m=0$; $\text{itr}=0$;

$Q=[]$; $Z=[]$; $S=[]$; $T=[]$;  

\While{$k<k_{\max}$ and $\text{itr}<\text{mx}$}{ Orthogonalize $t\leftarrow t-V_mV_m^*t$;

$\text{itr} \leftarrow \text{itr}+1$; $m=m+1$; $v_m=t/\|t\|$; $v^A_m=Av_m$; $v^B_m=Bv_m$; $w=\nu_0v^A_m+\mu_0v^B_m$;

Orthogonalize: $w\leftarrow w-Z_kZ_k^*w$;  $w\leftarrow w-W_{m-1}W_{m-1}^*w$; $w_m=w/\|w\|$;

$\displaystyle 
M^A\leftarrow  \begin{pmatrix}
    M^A & W_{m-1}^*v^A_m \\
    w_m^* V^A_{m-1} & w_m^*v^A_m 
  \end{pmatrix};
  \quad
 M^B\leftarrow   \begin{pmatrix}
    M^B & W_{m-1}^*v^B_m \\
    w_m^*V^B_{m-1} & w_m^* v_m^B 
  \end{pmatrix};
$ 

Compute the $QZ$ decomposition $M^AS^R=S^LT^A$, $M^BS^R=S^LT^B$, such
that
$|T^A_{i,i}/T^B_{i,i}-\tau|\leq|T^A_{i+1,i+1}/T^B_{i+1,i+1}-\tau|$
(the Rayleigh-Ritz step);

$u=Vs^R_1$; $p=W_js^L_1$; $u^A=V^As^R_1$; $u^B=V^Bs^R_1$; $\zeta=T^A_{1,1}$; $\eta=T^B_{1,1}$;

$r=\eta u^A-\zeta u^B$; $\wt{a}=Z^*u^B$; $\wt{b}=Z^*u^B$; $\wt{r}=r-Z(\eta\wt{a}-\zeta\wt{b})$;

\While{$\|\wt{r}\|<\epsilon$}{
$\displaystyle
R^A\leftarrow \begin{pmatrix}
    R^A & \wt{a} \\
    0^{\trans} & \zeta
  \end{pmatrix};
  \quad
  R^B\leftarrow  \begin{pmatrix}
    R^B & \wt{b} \\
    0^{\trans} & \eta
  \end{pmatrix};
$

$Q\leftarrow [Q,u]$; $Z\leftarrow [Z,p]$; $k\leftarrow k+1$; $m\leftarrow m-1$;

\If{$k=k_{\max}$}{
\Return $(Q,Z,R^A,R^B)$
}

\For{$i=1,\dots,m$}{
$v_i=Vs^R_{i+1}$; $v^A_{i}=V^As^R_{i+1}$; $v^B_i=V^Bs^R_{i+1}$;

$w_i=Ws^L_{i+1}$; $s^R_i=s^L_i=e_i$;
}

$M^A$, $M^B$ is the lower $m\times m$ block of $T^A$, $T^B$, respectively.

$u=u_1$; $p=w_1$; $u^A=v^A_1$; $u^B=v^B_1$; $\zeta=T^A_{1,1}$; $\eta=T^B_{1,1}$;

$r=\eta u^A-\zeta u^B$; $\wt{a}=Z^*u^A$; $\wt{b}=Z^*u^B$; $\wt{r}=r-Z(\eta\wt{a}-\zeta\wt{b})$;

}

\If{$m\geq m_{max}$}{
\For{$i=2,\dots,m_{\min}$}{
$v_i=Vs^R_i$; $v^A_i=V^As^R_i$; $v^B_i=V^Bs^R_i$; $w_i=Ws^L_i$;
}

$M^A$, $M^B$ is the leading $m_{\min}\times m_{\min}$ block of $T^A$, $T^B$, respectively.

$v_1=u$; $v^A_1=u^A$;$v^B_1=u^B$; $w_1=p$; $m=m_{\min}$.
}

$\wt{Q}=[Q,u]$; $\wt{Z}=[X,p]$;

(Approximately) solve the correction equation for $t\perp \wt{Q}$ using GMRES,
$(I-\wt{Z}\wt{Z}^*)(\eta A-\zeta B)(I-\wt{Q}\wt{Q}^*)t=-r$, where $r=\eta A u^A-\zeta B u^B$.

}

\end{algorithm2e}

\bibliographystyle{abbrv}
\bibliography{ref}

\end{document}